\documentclass{ws-procs11x85}
\usepackage{multicol,float} 

\begin{document}

\title{Detection of a relic X-ray jet in Cygnus A}

\author{K. C. Steenbrugge, K. M. Blundell}

\address{University of Oxford,\\
Oxford, OX1 3JP, UK\\
E-mail: kcs@astro.ox.ac.uk; kmb@astro.ox.ac.uk\\
}

\author{P. Duffy}

\address{School of Mathermatics, UCD,\\
Dublin, 4, Ireland\\
E-mail: peter.duffy@ucd.ie}

\begin{abstract}
We present a 200 ks {\it Chandra} ACIS-I image of Cygnus~A, and
discuss a long linear feature seen in its counterlobe. This feature
has a non-thermal spectrum and lies on the line connecting the
brighter hotspot on the approaching side and the nucleus. We therefore
conclude that this feature is (or was) a jet. However, the outer part of this
X-ray jet does not trace the current counterjet observed in radio. No
X-ray counterpart is
observed on the jet side. Using light-travel time effects we conclude
that this X-ray 50~kpc linear feature is a relic jet that contains enough
low-energy plasma ($\gamma \sim$ 10$^3$) to inverse-Compton scatter 
cosmic microwave background photons, producing emission in the X-rays.  
\end{abstract}


\bodymatter

\begin{multicols}{2}
\section{Introduction}\label{intro:sec1}

The Cygnus\,A cluster and galaxy (3C\,405) are one of the brightest
sources in the X-ray sky and have therefore been studied with every
major X-ray satellite. In this paper we take advantage of the high
spatial resolution of the {\it Chandra} satellite to study the linear
jet-like feature revealed in the X-ray image within the
counterlobe. The spatial resolution of the ACIS camera onboard {\it
Chandra} allows us to spatially resolve the jet, lobes and hotspots
from the central AGN.

Detection of X-ray photons that arise from inverse Compton
up-scattered Cosmic Microwave Background (ICCMB) photons mandate the
presence of relativistic particles with Lorentz factors of order
$10^3$ \cite{harris79}.  Such particles will have lower Lorentz
factors than ambient synchrotron-emitting particles radiating at the
typically-observed radio wavelengths, assuming the magnetic field
strengths in the lobes of radio galaxies are nT in size or lower.
Thus lower Lorentz factor particles can signal the presence of relic
(that is previously, but no longer detectably synchrotron emitting)
plasma \cite{erlund06,blundell06}.

With the physical size of Cygnus\,A being 130\,kpc, and a redshift of
0.05607 \cite{owen97}, (assuming a cosmology of $H_0$ = 73 km s$^{-1}$
Mpc$^{-1}$ and $\Omega_{\rm M} = 0.3$ and $\Omega_\Lambda$ = 0.7) the
light-travel time between opposite lobes exceeds $\cos \theta \times 4
\times 10^5$ years, where $\theta$ is the angle between the axis of
the radio source and our line-of-sight.  Since the light we observe
from opposite lobes is received at the same telescope time, this means
that an observer on Earth sees the nearer lobe at a more recent epoch
than the further lobe, which is seen at an earlier time (hence less
evolved) in the radio galaxy's history.

\section{{\it Chandra} Observations}

The details of the observations are listed in Table~1. The first 2
observations used the ACIS-S (Advanced CCD Imaging Spectrometer)
instrument and were only used in the spectral analysis. The remaining
eight observations analysed in this paper used the ACIS-I instrument
in the VFAINT mode, which gives a reduced background after
processing. All the data were obtained from the {\it Chandra} public
archive and reduced with the standard threads in CIAO 3.3, which
included the updated calibration database CALDB 3.2.2. The background
region was chosen from a low count rate region on the CCD containing
the image of Cygnus\,A.

\begin{figure}[H]
\begin{center}
 \resizebox{\hsize}{!}{\includegraphics[angle=0]{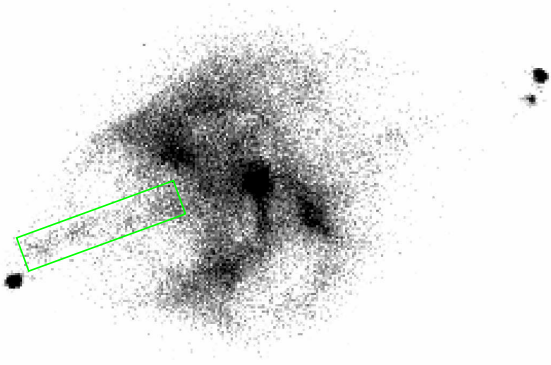}}
\caption{The 0.2$-$10\,keV ACIS-I image clearly showing the linear
  counterjet-like feature delineated by the green box and the
  non-detection of anything corresponding to this on the jet
  side.
  \label{fig:x-ray_cjet}} 
\end{center}
\end{figure}

We aligned the AGN core detected in the 2 $-$ 10 keV band of the
different observations using the reproject$\_$aspect thread and
then added the images using the merge$\_$all command in CIAO.  The
resulting image is shown in Fig.~\ref{fig:x-ray_cjet}. For extraction of
the spectra from different regions we used the specextract command
using the specific badpixel file for each observation. The quoted
errors on the X-ray luminosity and photon index is for $\Delta\chi^2$
= 2, the RMS of the $\Delta\chi^2$ distribution. We used the SPEX
\cite{kaastra02b} package for fitting the spectra.

\begin{table}[H]
\tbl{The list of the 8 observations of Cygnus\,A used in this paper.
   Listed are the date of the observation and the filtered exposure time.} 
{\begin{tabular}{|l|l|r|}\hline
    &   date     & exposure (ks) \\\hline
1   & 21 05 2000 & 34.72         \\
2   & 26 05 2000 & 10.17         \\
3   & 15 02 2005 & 25.80         \\
4   & 16 02 2005 & 51.09         \\
5   & 19 02 2005 & 25.44         \\
6   & 21 02 2005 & 6.96          \\
7   & 22 02 2005 & 23.48         \\
8   & 23 02 2005 & 23.05         \\
9   & 25 02 2005 & 16.04         \\
10  & 07 09 2005 & 29.65         \\\hline   
\end{tabular}}\label{tab:obs}
\end{table}

\section{Results}

\begin{figure}[H]
\begin{center}
 \resizebox{\hsize}{!}{\includegraphics[angle=-90]{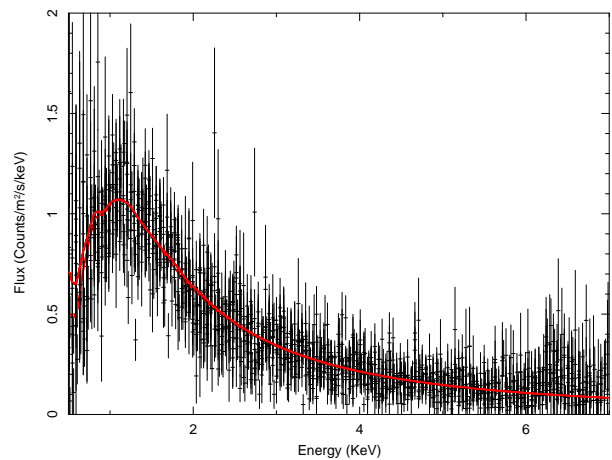}}
\caption{The 0.5$-$7\,keV flux spectrum for the 10 different
  observations. The slight excess around 6.4 keV is due to the thermal 
  emission. The slight different fit and larger spread in
  data points is a result of the different calibration between ACIS-I
  and ACIS-S.  
  \label{fig:spectrum}} 
\end{center}
\end{figure}

A linear counterjet-like feature is easily detected in the 0.2$-$10
keV image (see Fig.~\ref{fig:x-ray_cjet}). The width of the counterjet
is $\sim$5$^{\prime\prime}$, and is resolved in our {\it Chandra}
image. The width of the brightest knots observed in the 15-GHz image
are $\sim$2$^{\prime\prime}$.9 \cite{steenbrugge07}. The X-ray
counterjet feature is thus wider than its radio jets in Cygnus~A. The
green box in Fig.~\ref{fig:x-ray_cjet} indicates the region from which
we extracted the spectrum for this feature. Inevitably, this box will
contain some emission from the background thermal gas originating from
the cluster in which Cygnus~A is embedded. We rebinned the spectrum by
a factor of 3, and fitted the spectrum between 0.5$-$7 keV. The X-ray
spectrum of the counterjet is well fitted ($\chi^2$ = 1.1 for 1582
degrees of freedom) by a power-law with Galactic absorption
\cite{dickey90} of $3.5 \times 10^{25} {\rm m}^{-2}$.  The 2$-$10 keV
luminosity is (7.0 $\pm$ 0.12)$\times$10$^{35}$ W and the photon index
is 1.70 $\pm$ 0.02 (see Fig.~\ref{fig:spectrum}). The data are too poor
to constrain the thermal component from the cluster.

The brighter features in Cygnus A were studied \cite{wilson06} before, but
not the counterjet. This study convincingly showed that the higher
temperature gas is at the outer edge, i.e. the contact
discontinuity. The more centrally located gas is quite a bit cooler
ranging between 3.80 and 4.28 keV and can be explained as being due to
the jet break-out phase \cite{sutherland07a,sutherland07b}. The
location of the counterjet feature is inconsistent with it either
belonging to the gas heated by the jet break-out or the contact
discontinuity.

Fitting the counterjet feature with a thermal model gives a poorer
fit, namely $\chi^2$ = 1.2 (for the same degrees of freedom), for a
temperature of 6.3 keV and an emission measure of $5.59 \pm 0.13
\times 10^{71} {\rm m}^{-3}$.  From this emission measure, using $Y =
n_{\rm e}n_{\rm H}V$ we derive an electron density of $3.1 \times 10^5
{\rm m}^{-3}$. We assumed $n_{\rm e} = 2.1 n_{\rm H}$ and a volume of
$1.22 \times 10^{61}\,{\rm m}^{3}$.  The upper limit to the thermal
electron density \cite{dreher87} in the western lobe of Cygnus\,A is $4 \times
10^2\,{\rm m}^{-3}$ for an isotropic random magnetic
field; this could be higher by as much as two orders of magnitude if
the magnetic field has many reversals.  Even this extreme upper limit
still falls short of the necessary electron density. We thus conclude
that this feature is unlikely to be thermal.

The X-ray counterjet obeys the criteria \cite{bridle84} for jet
idenitification: its length is more than four times larger than its
width; it is separable at high spatial resolution from the surrounding
features; and it is aligned with the compact core. More precisely, the
counterjet lies on the same line that connects the brighter western
hotspot (i.e. the brighter hotspot on the jet side) and the
nucleus. Therefore, we conclude that the long linear feature
identified on the east of the source is indeed related to jet activity
and not part of the surrounding environment.

\subsection{Relic counterjet}

The inner part of the X-ray counterjet does overlay with the inner
part of the counterjet detected in the radio images, however it does
not make the 27$^{\circ}$31$^{\prime}$ bend observed in the 15-GHz
image \cite{steenbrugge07}; rather it extends along its original
direction until just north of the bright eastern radio hotspot. There is a
clear gap in emission between the end of the X-ray detected counterjet
and the radio hotspot. We conclude that the X-ray detected counterjet is a
relic for the following reasons: (i) it is extended transversely
compared to the radio counterjet and radio jet, (ii) it does not
overlay the outer radio counterjet, (iii) there is a gap in emission
between the observed counterjet and the hotspots.

The inner part of the relic counterjet overlaps with the inner part of
the current (radio) counterjet, and therefore we cannot constrain in this
region the X-ray luminosity potentially coming from the inner current
counterjet. However, for the outer X-ray counterjet there is a lack of
associated 15-GHz radio emission \cite{steenbrugge07}. This indicates
a lack of high Lorentz-factor particles in this relic counterjet
compared with those observed in active jets.

\section{Light-travel time effects and cooling}

The observation of X-ray emission from the counterjet side, and the
absence of such emission from the approaching side of the source,
places general constraints on the rate at which the X-ray emission
must fade. Assuming symmetrical conditions, this depends purely on
light-travel time arguments.  We consider approaching and receding
jets moving with speed $v=\beta_{\rm HS} c$ from the central source
and making an angle $\theta$ with the observer's line of sight. When
the approaching jet is observed to have extended out to a distance
$d$, so that $t_{\rm app}=d/v$ is the time since the source activity
began, the hotspot of the receding jet is observed at a younger age
given by:

\begin{equation}
t_{\rm rec}=\left({1-\beta_{\rm HS}\cos\theta}\over{1+\beta_{\rm
    HS}\cos\theta}\right)t_{\rm app}. 
\end{equation}

\noindent{Therefore, the difference in age between the observed jet
extremities, which must of course be imaged at the same ``telescope
time'', is}

\begin{eqnarray}
\displaystyle  \Delta t
=\frac{2\beta_{\rm HS}\cos\theta}{1 + \beta_{\rm HS}\cos\theta}t_{\rm
  app}. 
\end{eqnarray}

\begin{figure}[H]
\begin{center}
 \resizebox{\hsize}{!}{\includegraphics[angle=0]{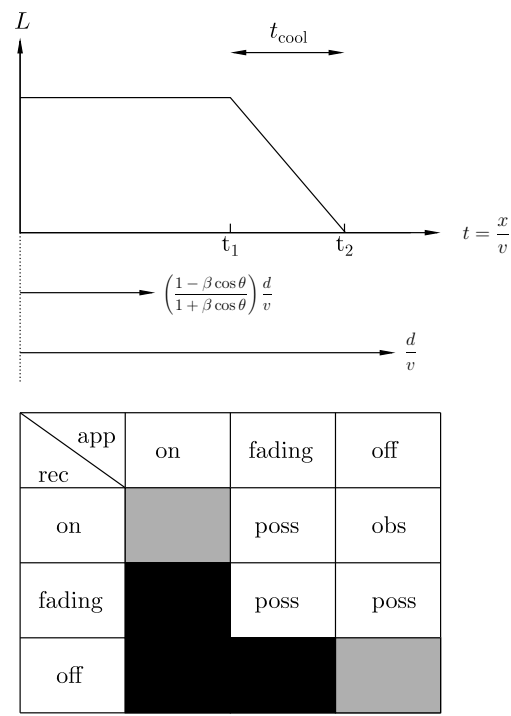}}
\caption{{\em Upper:} Schematic lightcurve for a fading
  jet. {\em Lower:} Illustrates the possible combinations for the
  fading jet and counterjet.  The black squares are not formally
  possible, because of light-travel time effects while those in grey
  are formally possible but excluded by observation.  The square
  labelled ``obs'' is the case we study in the text.
  \label{fig:lightcurve}} 
\end{center}
\end{figure}

What we actually observe is not just dependent on the kinematics of
light-travel time, but also on the evolution of the luminosity in the
jet material. To illustrate this point we consider the simple case
where the X-ray, ICCMB emission depends on the time since emission
from the central source in a manner qualitatively illustrated in
Fig.~\ref{fig:lightcurve}. 

In this picture the relic jet/counterjet plasma is moving at speed
$\beta_{\rm pl}$; this is very likely to be slower than the speed at
which the current jet or hotspots move otherwise we would not see any
relic plasma along the counterjet. When the cooling is sufficiently
rapid, then the cooling timescale $t_{\rm cool}<\Delta t$, and the
leading edge of the forward jet will be observed to cool and fade
before the far extremity of the counterjet is observed to cool. The
ratio of intensities for approaching and receding jets is dependent on
the flux density, $S_\nu\propto \nu^{-\alpha}$, with $\alpha\approx
0.7$.  We consider the flux ratio between equal volumes of plasma at
either extremity of the approaching and receding jets,

\begin{equation}
{S_{\rm app}\over S_{\rm rec}} = \left({1+\beta_{\rm pl}\cos\theta}\over
{1-\beta_{\rm pl}\cos\theta}\right)^{3+\alpha}
{L_{\rm app}(t_{\rm app})\over L_{\rm rec}(t_{\rm rec})}
\end{equation}
where we have ignored factors $O(\beta_{\rm pl}^2)$.  

Initially, when neither jet has started to fade, the forward jet will
be more luminous on the basis of size relative to the receding jet and
Doppler boosting. Then, as shown in Fig.~\ref{fig:lightcurve}, the
forward jet will be observed to fade first and the flux ratio will be
observed to decline, giving a much more prominent counterjet. We
measure the flux ratio for the relic jet over the relic counterjet in
the 2-10 keV band to be less than 0.25, as the upper limit to any jet
emission is about 4 times smaller than the measured counterjet
emission.  From the total observed length of the radio source of
130\,kpc, we calculate (assuming $\theta$ = 60$^{\circ}$) that the
light travel-time difference between the hotspots is 2 $\times$ 10$^5$
years.  Therefore the cooling time needs to be $\lesssim$ 10$^5$
years.  Thus, provided that the intrinsic luminosity of X-ray ICCMB
fades within a light-crossing time ($2 \times 10^5$ years), the system
will always evolve to a state where the receding jet has a greater
X-ray luminosity than the older, approaching counterpart.

\section{Conclusion}

We have analysed the X-ray counterjet revealed by the combined 200\,ks
{\it Chandra} ACIS-I image of Cygnus\,A. Its power-law spectrum, with photon
index of 1.7, as well as the upper limit to the thermal electron
density determined from radio polarimetry indicates
that the feature cannot be explained by thermal gas. It is therefore
likely to be emission from jet plasma having spectral index
0.7. Comparing the X-ray detected counterjet with the observed radio
counterjet in the 5-GHz and 15-GHz radio images, we conclude that the
counterjet detected in X-rays is a relic jet. This conclusion was
reached from the following observations: (i) the curvature of the
outer parts of the X-ray counterjet is significantly different from that of
the current radio counterjet; (ii) this feature lacks any directly associated
radio emission implying a lack of high energy synchrotron particles;
and (iii) the width of the X-ray counterjet is significantly broader
than the radio jet or counterjet implying expansion.

\section*{Acknowledgments}
The authors would like to thank St John's College, Oxford; the Royal
Society and the Royal Irish Academy.

\bibliographystyle{ws-procs11x85}
\bibliography{references}

\end{multicols}
\end{document}